\documentclass[]{aa}  

\usepackage{graphicx}
\usepackage[varg]{txfonts}
\usepackage{natbib}

\newcommand{\rsoph}{RS~Oph}
\newcommand{\ha}{H$\alpha$}

\newcommand{\kms}{\,km\,s$^{-1}$}
\newcommand{\hi}{H\,{\small I}}
\newcommand{\hei}{He\,{\small I}}
\newcommand{\heii}{He\,{\small II}}  

\def\FeII{\ion{Fe}{II}}        

\begin{document}

\title{Mass outflow from the symbiotic binary RS~Oph\\ during its 2021 outburst\thanks{Based on                      data collected with 2-m RCC telescope at Rozhen National Astronomical Observatory}
}  
\subtitle{}     
\author{N.~A.~Tomov
        \inst{\ref{inst1}}\fnmsep\thanks{$\dagger${\rm ~Deceased}}
        \and
        M.~T.~Tomova
        \inst{\ref{inst1}}\fnmsep\thanks{email: mtomova@astro.bas.bg}
        \and
        K.~A.~Stoyanov
        \inst{1}
        \and
        T.~R.~Bonev
        \inst{1}
        \and
        R.~K.~Zamanov
        \inst{1}
        \and
        I.~Kh.~Iliev
        \inst{1}
        \and
        Ya.~M.~Nikolov
        \inst{1}
        \and
        D.~Marchev
        \inst{\ref{inst2}}
        \and
        D.~V.~Bisikalo
        \inst{\ref{inst3}}
        \and
        P.~V.~Kaygorodov
        \inst{\ref{inst3}}
}

\institute{Institute of Astronomy and NAO, Bulgarian Academy of 
           Sciences, Tsarigradsko Shausse 72, 1784, Sofia, Bulgaria \label{inst1}
       \and
           Department of Physics and Astronomy, Shumen University Episkop Konstantin
                   Preslavski, 115 Universitetska Str., 9700 Shumen, Bulgaria \label{inst2}
       \and
           Institute of Astronomy of the Russian Academy of Sciences, 
           48 Pyatnitskaya Str., 119017 Moscow, Russia \label{inst3}
}

\date{Received / Accepted}

\abstract
  { \object{\rsoph} is a symbiotic recurrent nova containing a massive white dwarf with heavy mass loss during activity. In August 2021, it underwent its seventh optical eruption since the end of the 19th century.}
        {The goal of this work is to analyse the structure of the outflows from the outbursting object.}
   {Based on broad-band $U$, $B$, $V$, $R_{\rm C}$, and $I_{\rm C}$ photometry and high-resolution {\ha} spectroscopy obtained at days 11--15 of the outburst, we derived some parameters of the system's components and outflows and their changes during our observation.}  
   {The effective temperature of a warm shell (pseudophotosphere) produced by the ejected material and occulting the hot component of the system was $T_{\rm eff}\!=\!15\,000\pm1\,000$~K and  the electron temperature of the nebula was $T_{\rm e}\!=\!17\,000\pm3000$~K throughout the observations. The effective radius of the pseudophotosphere was $R_{\rm eff}\!=\!13.3\pm2.0$~R$_{\sun}$ and the emission measure of the nebula $EM\!=\!(9.50\pm0.59)\,$10$^{61}$~cm$^{-3}$ for day 11 and $R_{\rm eff}\!=\!10.3\pm1.6$~R$_{\sun}$ and $EM\!=\!(5.60\pm0.35)\,$10$^{61}$~cm$^{-3}$ for day 15. To provide this emission measure, the bolometric luminosity of the outbursting object must exceed its Eddington limit. The mass-loss rate of the outbursting object through its wind is much greater than through its streams. The total rate (from wind + streams) was less than $(4-5)\,$ 10$^{-5}$ (d/1.6kpc)$^{3/2}$ M$_{\sun}$\,yr$^{-1}$. The streams are not highly collimated. 
    Their mean outflowing velocities are $\upsilon_{b}\!=\!-3\,680\pm60$\kms\ for the approaching stream and $\upsilon_{r}\!=\!3\,520\pm50$\kms\ for the receding one if the orbit inclination is 50\degr.   
   }   
   {}


\authorrunning{Tomov et al.}
\titlerunning{Mass outflow from RS~Oph }

\maketitle
%
%

\section{Introduction}

   Symbiotic stars are detached binaries consisting of a normal cool giant of spectral type G, K, M, or Mira and a hot compact object accreting mass from the wind of the giant. As a result of the accretion, symbiotic stars evolve through phases of quiescence and activity. The profiles of the spectral lines of some of these binaries have high-velocity satellite components indicating mass loss through bipolar collimated outflows. However, most of these systems show indications of collimated outflows only during active phases, when in addition to the satellite components, their profiles contain wind components; for example P~Cyg absorptions and/or broad emissions formed by an optically thin high-velocity wind (e.g. \object{MWC~560} \citep{TK97}; \object{Hen~3-1341} \citep{toma00}; \object{Z~And} \citep{TTB07}; RS~Oph \citep{sk+08}; and \object{BF~Cyg} \citep{sk+13}).
 Broad emissions are determined by the kinematics of the emitting gas and in some cases their width 
    is a measure of
 the speed of a shock wave generated by the ejecta \citep{sokoloski+06}.
 The \ha\ line of the symbiotic stars is emitted in all parts of their circumbinary envelope and its profile contains several  components during activity.
   
   The system \rsoph\ is a prototype of the recurrent symbiotic novae, and consists of an M2\,III giant \citep{zam+18} and a white dwarf with a mass close to the Chandrasekhar limit \citep[][and references therein]{bode87} with an orbital period of 456$^d$ \citep{f+00,br+09}. 
   Seven optical outbursts of \rsoph\, with light maxima in 1898, 1933, 1958, 1967, 1985, 2006, and 2021 have been recorded.

A typical feature of the system \rsoph\ during its previous outbursts is the shape of the remnant from the nova explosion, which is far from spherical and is more close to a bipolar nebula. During the previous 2006 outburst, radio data were obtained with MERLIN, VLA, VLBA, and EVN arrays \citep{o'brien+06}. The data obtained at days 13$\div$49 after the beginning of the outburst reveal several components of the radio image that appear at different times.  This behaviour is thought to result from an expanding shock wave, which sweeps through the wind of the cool giant creating a high-temperature region. The components of the radio image were modelled with a bipolar shock-heated shell whose axis is perpendicular to the orbital plane. The hypothesis of the outward-moving shock wave was confirmed by the results of X-ray observations \citep{sokoloski+06}, which allowed the conditions within the ejecta to be diagnosed. The flux in the energy range 2 -- 20 keV was fitted with a single-temperature thermal free-free emission, \FeII\ emission lines, and absorption by intervening material. During the first three weeks of the outburst, the flux faded strongly, and its high-energy part faded the most, indicating a drop in temperature. The rapid fading of the X-ray flux suggests that the nova explosion deviates from a spherically symmetric blast wave. The system \rsoph\ was observed in the X-ray domain several years after its 2006 outburst as well. In 2009 and 2011, the Chandra Observatory detected an extended emission consistent with a bipolar flow seen from the system in opening angles of about 70$^{\circ}$ \citep{montez+22}. The lengths of both lobes grew and in 2011 the lengths reached 2.0 arcsec projected on the sky. No evidence of cooling of the ejected gas was found and on this basis it was concluded that it had expanded freely, driven by some mechanism of collimation away from the system. It was also concluded that the ejected gas moved into some cavity formed by the 1985 outburst. The circumbinary medium in \rsoph\ was modelled by \citet{booth+16}. One of the results of this latter study is that the mass transferred to the white dwarf concentrates towards the orbital plane and forms an accretion disc that extends to the boundary of the Roche lobe. The spherical ejecta of the outbursting dwarf interact with the disc and produce a highly bipolar nebular structure with a dense equatorial ring.
   
At the very beginning of the previous 2006 outburst, the \ha\ profile of \rsoph\ comprised three components: a weak narrow central double peak emission, a very intensive triangular broad component with 
 a full width at zero intensity (FWZI) of about
 7600~\kms\ emitted in an optically thin stellar wind from the outbursting object, and a blueshifted absorption with a velocity of about 4250\kms. After that, the line changed, acquiring a central intensive emission with a triangular shape again and FWZI $\sim 3500$\kms, which was interpreted as a slowly expanding equatorial ring-like structure and satellite emissions on both sides of the central one with velocity position of $\sim 2500$\kms, indicating bipolar collimated outflow. After the first month of the outburst, the satellite emissions disappeared and a high-velocity wind was observed in the \ha\ wings only \citep{sk+08}. Satellite emission components at the same velocity position in the lines Pa$\beta$ and Br$\gamma$ were observed during the first month of this outburst by \citet{ban+} as well. In addition, satellite emissions of the lines of \hei\ and \heii\ with much lower velocities of about 200\kms\ were observed by \citet{Iijima09} and \citet{br+09} in March and April 2006.


\begin{table*}[!htb]
\caption{Journal of observations of \rsoph.}
\label{tab:1}
\begin{center}
\begin{tabular}{lccclccccc}
\hline\hline
    \noalign{\smallskip}
Date & \multicolumn{3}{c}{Spectral Data} & & \multicolumn{5}{c}{Photometric Data\tablefootmark{a}} \\
       \cline{2-4} \cline{6-10}
    \noalign{\smallskip}
 & UT start & Exp. [s] & HJD$-$2459000\tablefootmark{b} & & $F_U$ & $F_B$ & $F_V$ & $F_{R_{\mathrm C}}$ & $F_{I_{\mathrm C}}$ \\
    \noalign{\smallskip}
\hline
    \noalign{\smallskip}
2021 Aug 19 & 19$^{\rm h}$56$^{\rm m}$02$^{\rm s}$ & 180 & 446.340 & & & 31.256 & 21.348 & 20.765 & 14.935 \\
                                                & 20 01 03 & 300 & & & & & & & \\
2021 Aug 20 & 19 14 04 & 120 & 447.316 & & & 23.709 & 15.466 & 16.494 & 12.422 \\
                                                & 19 17 56 & 300 & & & & & & & \\
                                                & 19 25 13 & 300 & & & & & & & \\
                                                & 19 31 29 & 300 & & & & & & & \\
2021 Aug 21 & 18 38 39 & 300 & 448.304 & & & 22.642 & 13.470 & 15.043 & 10.819 \\
                                                & 18 45 25 & 600 & & & & & & & \\
                                                & 18 57 37 & 600 & & & & & & & \\
                                                & 19 10 11 & 600 & & & & & & & \\
                                                & 19 24 09 & 500 & & & & & & & \\
                                                & 19 33 48 & 400 & & & & & & & \\
2021 Aug 22 & 18 55 28 & 300 & 449.305 & & & 19.720 & 12.285 & 13.719 & ~9.867 \\
                                                & 19 01 44 & 300 & & & & & & & \\
                                                & 19 08 03 & 300 & & & & & & & \\
                                                & 19 13 48 & 300 & & & & & & & \\
                                                & 19 19 42 & 300 & & & & & & & \\
2021 Aug 23 & 19 31 01 & 300 & 450.334 & & 54.646 & 17.985 & 12.285 & 13.102 & ~9.423 \\
                                                & 19 46 49 & 300 & & & & & & & \\
                                                & 19 54 19 & 300 & & & & & & & \\
                                                & 20 00 25 & 300 & & & & & & & \\
                                                & 20 06 10 & 300 & & & & & & & \\
\hline
\end{tabular}
\end{center}
\tablefoot{
\tablefoottext{a}{Dereddened fluxes based on the light curves of AAVSO, in units of $10^{-12}$ erg\,cm$^{-2}$\,s$^{-1}$\,\AA$^{-1}$.}
\tablefoottext{b}{A middle moment in time of all exposures during one night.}
}

\end{table*}

   The present 2021 outburst of \rsoph\ began on August 8.93 \citep{geary21} and was observed over the whole electromagnetic domain. High-resolution optical spectroscopy during its first 2--3 days was obtained by \citet{MV21a}, \citet{Mik+21} and \citet{taguchi+21}. One excellent atlas showing the evolution of the optical spectrum of \rsoph\ from August 9 to 26 and containing data from every night in this period was presented by \citet{MV21b}. At the beginning of the outburst, the \ha\ profile comprised an intensive narrow central emission with a sharp absorption on its blue side, an intensive triangular broad component with FWZI $\sim 6\,000$\kms\ produced by the ejecta, and a deep P~Cyg absorption at $\sim$3\,000--4\,000\kms, which disappeared after August 13 . The central emission decreased strongly during the observation. In the first two days, the width of the broad emission increased and its FWZI reached about 7\,000\kms, keeping this value until the end of observation. However, the 
   full width at half maximum (FWHM)
    of the line began to decrease after August 12 and the profile acquired a shape more closely resembling a bell than a triangle. In addition, the \ha\ line had weak emission bumps on its wings, which could be the signature of a bipolar outflow; these were visible from August 12 to 26. 

We observed the \ha\ line in several nights starting 11 days after the beginning of the outburst. Our main aim is to investigate the structure of the outflowing material from the outbursting object on the basis of the \ha\ profile. One complementary task is to estimate its mass-loss rate.

\section{Observations and data reduction}

The region of the H$\alpha$ line of \rsoph\ was observed with the Andor Newton CCD camera mounted on the Coude spectrograph of the 2m Ritchey-Chretien-Coude (RCC) telescope of the National Astronomical Observatory Rozhen, Bulgaria. The observed wavelength window was 225 \AA\ and the spectral resolution was 0.11~\AA\,px$^{-1}$. Several spectra were taken every night from August 19 to 23. The list of observations is presented in Table~\ref{tab:1}.
Some of these spectra were used in the work of \citet{zam+22}.
The spectra from each particular night were added together to improve their signal-to-noise ratio (S/N). The data were reduced in the standard way using the IRAF package\footnote {The IRAF package is distributed by the National Optical Astronomy Observatories,
which is operated by the Association of Universities for Research in Astronomy, Inc., under contract with the National Science Foundation.}. 

To build a spectral energy distribution of the system we used average $U$, $B$, $V$, $R_{\rm C}$, and $I_{\rm C}$ photometric estimates from the light curves of the AAVSO observers taken during our observations.
The energy fluxes were calculated using data from
Bessell (1979) for a zero-magnitude star. The continuum flux at the position of H$\alpha$ was obtained using linear interpolation of the $R_{\rm C}$ and $I_{\rm C}$ continuum fluxes. The H$\alpha$  flux was obtained from the equivalent width and continuum flux at the position of this line. The $BVR_{\rm C}I_{\rm C}$ fluxes were corrected for the strong emission lines of \rsoph\ by means of low-resolution spectra from the Astronomical Ring for Access to Spectroscopy database\footnote {http://aras-database.github.io/database/symbiotics.html} \citep[ARAS;][]{teyssier19} also taken during our observations. The ARAS spectra are of different S/N. We measured the equivalent widths of eight lines in $B$ band, eight in $V$ band, six in $R_{\rm C}$ band, and 3 in $I_{\rm C}$ band. In addition to these lines, \rsoph\ has other weak emission lines that are not measurable and whose total contribution is smaller than the uncertainty of the continuum fluxes. We calculated a correction for every night, but as it changed within the limits of our accuracy, we took its arithmetical mean for all spectra. Its mean values are $41.8\pm2.2$\,\%, $25.4\pm1.8$\,\%, $53.4\pm2.4$\,\%, and $15.0\pm1.7$\,\% for the photometric bands $B$, $V$, $R_{\rm C}$, and $I_{\rm C}$. However, the range of the ARAS spectra does not include the photometric band $U$ and moreover these spectra are too noisy in the region  of the Balmer jump. Nevertheless, we used a $U$ flux available for August~23 without these corrections but dereddened. The $UBVR_{\rm C}I_{\rm C}$ fluxes were obtained with an uncertainty of 4.6\,\% and are listed in  Table~\ref{tab:1}. 
The continuum and \ha\ fluxes were dereddened with use of $E(B-V)\!=\!0.69$ \citep{zam+18} and the extinction curves of \citet{Cardelli}.

\section{Continuum energy distribution}

To examine the outflow structure of the outbursting compact object, we require an estimate of the inner boundary of the outflows. The spectral energy distribution of the system {\rsoph} shows that the observed photosphere (pseudophotosphere) of the outbursting object at the time of maximal light in 2006 and several days later had an enormous effective radius of 160--200~R$_{\sun}$ and a low temperature of about 6\,000--9\,000~K. After the seventh day of the outburst, the radius decreased to 1--2~R$_{\sun}$ and the temperature increased to about 110\,000~K \citep{sk15a}. This is why it was also necessary to build the energy distribution at the time of our spectral observations in order to determine the effective radius of the pseudophotosphere. We were able to provide ourselves with $UBVR_{\rm C}I_{\rm C}$ photometry only and were not able to find any UV flux at this time. The duration of our observations was five days (Table~\ref{tab:1}) and the brightness of {\rsoph} changes  slightly over one day. If we build the energy distribution for every day, the distributions will differ slightly but within the limits of our accuracy. This is why we only built the energy distribution for the first and last day of our observations. 


\begin{figure}

\includegraphics
{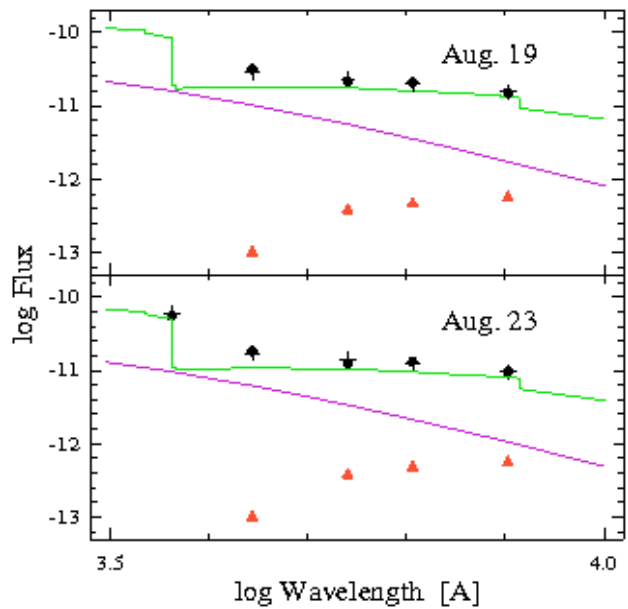}
 \caption{Continuum energy distribution of RS Oph in the $UBVR_{\rm C}I_{\rm C}$ range. The lines designate the black body and nebular continua and the triangles show the fluxes of the giant. The crosses designate the total fluxes and the points show the observed fluxes. 
}
 \label{sed}
\end{figure}


\begin{table*}
\caption{Continuum fluxes of the system's components in units of $10^{-12}$ erg\,cm$^{-2}$\,s$^{-1}$\,\AA$^{-1}$. The uncertainties of the observed values are in the same units.}
\label{tab:2}
\begin{center}
\begin{tabular}{@{}lrrrrcrrrrr@{}}
\hline\hline
    \noalign{\smallskip}
SC & \multicolumn{4}{c}{2021 Aug 19} & & \multicolumn{5}{c}{2021 Aug 23} \\
       \cline{2-5} \cline{7-11}
    \noalign{\smallskip}

 & $F_B$ & $F_V$ & $F_{R_{\mathrm C}}$ & $F_{I_{\mathrm C}}$ & & $F_U$ &
         $F_B$ & $F_V$ & $F_{R_{\mathrm C}}$ & $F_{I_{\mathrm C}}$ \\
          \noalign{\smallskip}
\hline
    \noalign{\smallskip}
Cool & 0.100 & 0.380 & 0.468 & 0.562 & & 0.0 &
                         0.100 & 0.380 & 0.468 & 0.562 \\
Hot  & 10.165 & 5.538 & 3.526 & 1.733 & & 9.497 & 6.097 & 3.321 & 2.115 & 1.040 \\
Nebula & 18.042 & 17.398 & 15.937 & 13.317 & & 49.129 & 10.635 & 10.256 & 9.394 & 7.850 \\
TF & 28.307 & 23.316 & 19.931 & 15.612 & & 58.676 & 16.832 & 13.957 & 11.977 & 9.452 \\
OF & 31.256 & 21.348 & 20.765 & 14.935 & & 54.646 & 17.985 & 12.285 & 13.102 & 9.423 \\
   & $\pm1.440$ & $\pm0.984$ & $\pm0.956$ & $\pm0.688$ & & $\pm2.514$ &
     $\pm0.828$ & $\pm0.566$ & $\pm0.604$ & $\pm0.434$ \\
r  & $-$9.4 & 9.2 & $-$4.0 & 4.5 & & 7.3 & $-$6.4 & 13.6 & $-$8.6 & 0.3 \\
\hline
    \noalign{\smallskip}
\end{tabular}
\end{center}
\tablefoot{
TF is total flux (TF~=~Cool+~Hot+~Nebula); OF is observed flux; r~=~(TF$-$~OF)/~OF in \%.
}
\end{table*}


\begin{table}
\caption{Product r~=~(TF$-$OF)/OF in \% for August 23 for a number of continuum distributions obtained with a relevant set of electron temperatures for several effective temperatures.}
\label{tab:3}
\begin{center}
\begin{tabular}{@{}rrrrrrrr@{}}
\hline\hline
    \noalign{\smallskip}
$T_{\rm eff}$(K) & $T_{\rm e}$(K) & \multicolumn{5}{c}{r (\%)} & $\sum{r}$ \\
       \cline{3-7} 
    \noalign{\smallskip}

 & & $F_U$ & $F_B$ & $F_V$ & $F_{R_{\mathrm C}}$ & $F_{I_{\mathrm C}}$ & \\
          \noalign{\smallskip}
\hline
    \noalign{\smallskip}
                & 17\,000 & 16.2 & 4.4 & 17.9 & $-$7.3 & 0.3 & 31.5 \\
25\,000 & 20\,000 & 6.1 & 18.1 & 28.4 & $-$2.7 & 0.3 & 50.2 \\
        & 25\,000 & 4.1 & 21.5 & 30.5 & $-$1.8 & 0.3 & 54.6 \\
        & 30\,000 & 1.9 & 25.5 & 33.0 & $-$0.6 & 0.3 & 60.1 \\
        & 35\,000 & $-$0.6 & 29.8 & 36.0 & 0.8 & 0.3 & 66.3 \\
        & 40\,000 & $-$3.3 & 34.7 & 39.2 & 2.2 & 0.3 & 73.1 \\
\hline
    \noalign{\smallskip}    
        & 17\,000 & 24.5 & 13.4 & 21.2 & $-$6.3 & 0.3 & 53.1 \\   
50\,000 & 20\,000 & 14.4 & 27.1 & 31.7 & $-$1.8 & 0.3 & 71.7 \\
        & 25\,000 & 12.4 & 32.0 & 33.8 & $-$0.8 & 0.3 & 77.7 \\
        & 30\,000 & 10.2 & 34.5 & 36.3 & 0.4 & 0.3 & 81.7 \\
        & 35\,000 & 7.7 & 38.8 & 39.4 & 1.7 & 0.3 & 87.9 \\
        & 40\,000 & 5.0 & 43.7 & 42.5 & 3.2 & 0.3 & 94.7 \\
\hline
    \noalign{\smallskip}        
                 & 17\,000 & 29.0 & 18.0 & 22.8 & $-$5.8 & 0.3 & 64.3 \\
100\,000 & 20\,000 & 18.9 & 31.8 & 33.3 & $-$1.3 & 0.3 & 83.0 \\
         & 25\,000 & 17.0 & 35.1 & 35.4 & $-$0.3 & 0.3 & 87.5 \\
         & 30\,000 & 14.7 & 39.2 & 37.9 & 0.9 & 0.3 & 93.0 \\
         & 35\,000 & 12.2 & 43.5 & 41.0 & 2.2 & 0.3 & 99.2 \\
         & 40\,000 & 9.5 & 48.3 & 44.1 & 3.7 & 0.3 & 105.9 \\
\hline
    \noalign{\smallskip}         
15\,000 & 17\,000 & 7.3 & $-$6.4 & 13.6 & $-$8.6 & 0.3 & 6.2 \\
\hline
    \noalign{\smallskip}
\end{tabular}
\end{center}

\end{table}

To calculate the fluxes of the pseudophotosphere and circumbinary nebula, we need to know the fluxes of the cool giant. Assuming that the giant does not change \citep{sk15a,sk15b}, we subtracted the fluxes of the giant obtained by \citet{sk15a} from the observed fluxes and approximated their residuals separately with a Planck function and nebular continuum determined by recombinations and free-free transitions.
 The Planck continuum is determined by the effective temperature and radius of the observed photosphere (pseudophotosphere) of the outbursting object.
 The nebular continuum is determined by the electron temperature and emission measure of the nebula.
  We built a number of model continua, varying these four parameters, and estimated their values at which the residual between the model fluxes of the system and the observed ones is minimal.
 As required by our treatment, we accepted  a helium abundance of 0.1 \citep{VN94} in the nebula. In order to model the nebular continuum, we had to also determine the dominant state of ionisation of helium.
The data of \citet{MV21b} show that the lines of {\heii} are not present in the spectrum of {\rsoph} during the first 18 days of the outburst. That is why we accepted that the dominant state of ionisation of helium in the nebula of {\rsoph} is {\heii} and its continuum is produced by  {\hi} and {\hei}. We used continuum emission coefficients of hydrogen from the paper of \citet{ferland80}
 and those of helium from the book of \citet{Pottash}. We accepted a distance to the system of 1.6 kpc, which is the value used in a very comprehensive analysis of the spectral energy distribution by \citet{sk15a,sk15b}.
  We calculated the parameter $\chi^2$ using the chi-square method for every one of our solutions, and for the best ones we obtained $\chi^2\!=\!0.52$ for August 19 and $\chi^2\!=\!0.70$ for August 23.
The approximation of the $UBVR_{\rm C}I_{\rm C}$ fluxes of {\rsoph} showed that, for the period of our observation, the effective temperature of the pseudophotosphere and  the electron temperature of the nebula have not changed; the former was 
$T_{\rm eff}\!=\!15\,000\pm1\,000$~K, and the latter $T_{\rm e}\!=\!17\,000\pm3000$~K.
The error of the electron temperature was estimated from a comparison of nebular fluxes with different temperatures. This error was calculated so that the residual of the fluxes with different temperatures is not smaller than the uncertainty of the observed fluxes. The error of the effective temperature of the outbursting object was estimated in the same way.
 We obtained an effective radius for the pseudophotosphere of $R_{\rm eff}\!=\!(13.3\pm2.0)$(d/1.6kpc)~R$_{\sun}$ and an emission measure for the nebula of $EM\!=\!(9.50\pm0.59)$\,10$^{61}$(d/1.6kpc)$^{2}$~cm$^{-3}$ for August 19 and $R_{\rm eff}\!=\!(10.3\pm1.6)$(d/1.6kpc)~R$_{\sun}$ and $EM\!=\!(5.60\pm0.35)$\,10$^{61}$(d/1.6kpc)$^{2}$~cm$^{-3}$ for August 23.
The errors of these parameters were obtained from the errors of the temperature and observed fluxes. 
The energy fluxes of the components of the system (SC) are listed in Table~\ref{tab:2} where the total fluxes of the system are compared with the observed fluxes. The continuum energy distribution is shown in Fig.~\ref{sed}.

We obtained the parameters of the pseudophotosphere using only a narrow $UBVR_{\rm C}I_{\rm C}$ range. 
We 
 tried to approximate the fluxes for August 23 with black-body continua with temperatures of 25\,000~K, 50\,000~K, and 100\,000~K and at the same time each of these three continua was used together with nebular continua with temperatures of 17\,000~K, as well as 20\,000~K to 40\,000~K, in steps of 5\,000~K (Table~\ref{tab:3}). Table 3 shows the same product r which persists in the last row of Table~\ref{tab:2} for each continuum distribution. The last row of Table~\ref{tab:2} is added for comparison.  If we approximate the observed flux at the position of the $I_{\rm C}$ band in each of these examples, the model flux exceeds the observed one at the position of the shorter wavelength bands because of an increase in the nebular emission 
towards the short wavelengths   coupled with a much steeper increase in the black-body emission. We should remember that the nebular emission produced by recombinations and free-free transitions increases more steeply than our observed fluxes from $I_{\rm C}$ band towards the shorter wavelength bands at all electron temperatures in the range 10\,000 -- 40\,000~K. This is why in our case it was possible to approximate the observed fluxes of \rsoph\ with only a black-body emission with a temperature not higher than about 15\,000~K and a nebular emission with a temperature of about 17\,000~K.
The spectrum of a black-body with such a temperature has a maximal intensity at $\lambda\lambda$ 1932~\AA. Its intensity at the wavelength of the $B$ band is only smaller  by a factor of 3.4. In this way, such a black-body emits an appreciable part of its spectrum in the $UBVR_{\rm C}I_{\rm C}$ range, which gives us reason to conclude that the parameters obtained are reliable. 
 
Using the determined quantities, we are able to compare the parameters of the pseudophotosphere and the emission measure of the nebula. The condition for ionisation equilibrium between the number of ionising photons (Lyman photon luminosity) and the rate of recombination in the nebula is 

\begin{equation}
        \mu = \frac{L_{\rm ph}}
        {\alpha_{\rm B}({\rm H},T_{\rm e})EM}\,\,,
\end{equation}

\noindent
where $\alpha_{\rm B}$(H,$T_{\rm e}$) is the recombination coefficient to all but the ground state of hydrogen (Case B). If $\mu \gid 1,$  only radiative ionisation is realised. The equality is fulfilled when all photons are absorbed in the nebula. If $\mu < 1,$ both kinds of ionisation are realised, radiative and shock. Their ratio is $(1-\mu) / \mu,$ where $\mu$ denotes the nebular continuum flux produced by the radiative ionisation and  $1-\mu $ is the nebular continuum flux produced by the shock ionisation. In some cases of distribution of the circumbinary gas, some of the ionising photons can leave the nebula. In these cases, $(1-\mu) / \mu$ is a lower limit of the ratio of the continua produced by shock and radiative ionisations.
 In our calculations, we used $\alpha_{\rm B}$(H,$T_{\rm e}$)\!=\!1.67\,$10^{-13}$~cm$^{3}$s$^{-1}$ and the function giving the number of ionising photons for hydrogen $G_{0}\!=\!0.1$ from the paper of \citet{NV87}. For August 19, we obtained $L_{\rm ph}\!=\!(2.30\pm0.93)\,10^{47}$~phot\,s$^{-1}$ and 
$\alpha_{\rm B}$(H,$T_{\rm e})EM\!=\!(1.74\pm0.11)$\,10$^{49}$~s$^{-1}$ and for August 23 $L_{\rm ph}\!=\!(1.38\pm0.57)\,10^{47}$~phot\,s$^{-1}$ and $\alpha_{\rm B}$(H,$T_{\rm e})EM\!=\!(1.03\pm0.06)$\,10$^{49}$~s$^{-1}$. In both cases, $L_{\rm ph} << \alpha$(H,$T_{\rm e})EM$ 
which means that both kinds of ionisation are realised. On the other hand, the X-ray observations show that only small areas in the nebula are heated by shock and have a high temperature of $10^6 \div 10^7$~K and an emission measure of 10$^{58}$~cm$^{-3}$ \citep{sokoloski+06}. We obtained a low effective temperature of 17\,000~K and an emission measure of about 10$^{62}$~cm$^{-3}$, which exceeds  the emission measure of the shock-heated areas by a factor of four orders of magnitude. We therefore prefer the supposition that the circumbinary nebula in \rsoph\ is mainly radiatively ionised; in other words, the observed rate of 
 recombination can be produced by a hot ionising star with $L_{\rm ph}\gid 1.74\,10^{49}$~phot\,s$^{-1}$ and $L_{\rm ph}\gid 1.03\,10^{49}$~phot\,s$^{-1}$ in the two cases. The star is occulted by its pseudophotosphere and ionises the nebula up and down, which means that this pseudophotosphere should be disc-shaped. Such a shell can result from interaction of the outflowing material with an accretion disc (see the Sect.~\ref{analysis ha}). This is why for the inner boundary of the outflows (ejecta), we should not accept the radius of the pseudophotosphere (shell) but rather a radius  close to that of the white dwarf in the quiescent state of the system. A white dwarf with a mass of 1.2$-$1.4~M$_{\sun}$ has a radius of 0.006$-$0.002~R$_{\sun}$ \citep{magano+17}. The effective radius of the underlying (occulted by the shell) outbursting component is not likely to be equal to the radius of the white dwarf because of expansion. \citet{sk15b} came to the conclusion that the temperature of the outbursting object 100 days after the beginning of the 2006 outburst is higher than 100\,000~K. 
 We set the temperature and effective radius of the underlying outbursting object in such a way that
  its Lyman luminosity is equal to or a little greater than the number of recombinations in the nebula for the two dates.
 If we accept  a temperature of 150\,000~K and effective radius of 0.9~R$_{\sun}$ for August 19, the Lyman luminosity will amount to $L_{\rm ph}\!=\!2.12\,10^{49} \ga 1.74\,10^{49}$~phot\,s$^{-1}$. If we accept the same temperature and an effective radius of 0.7~R$_{\sun}$  for August 23, the Lyman luminosity will amount to $L_{\rm ph}\!=\!1.28\,10^{49} \ga 1.03\,10^{49}$~phot\,s$^{-1}$. This is why for the inner boundary of the outflows during our observations, we will take a radius for each date interpolated between 0.9~R$_{\sun}$ and 0.7~R$_{\sun}$. 

\section{Analysis of the {\ha}  profile and mass-loss rate}\label{analysis ha}


\begin{figure*}

\includegraphics[width=\hsize]
{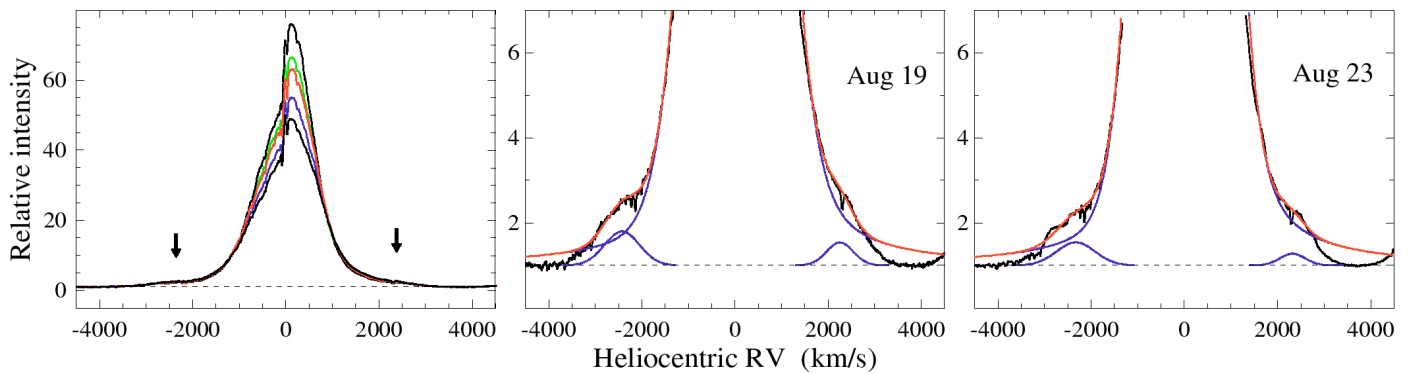}
 \caption{\ha\ line. {\it Left panel:} Evolution of \ha\ line. The spectrum of lowest intensity relates to August 19. {\it Middle and right panels:} Area of the wings of the line where satellite components are seen. Approximating curves are also shown: blue colour shows line components, and the red colour shows the resulting curve approximating the observed spectrum.
}
 \label{ha_all}
\end{figure*}

At the time of our observations, the central narrow emission was very weak and we considered only the very intensive broad component of {\ha}. This component had an appreciated asymmetry and very broad low-intensity wings reaching  $\pm$3\,500\kms\ from the centre of the line (Fig.~\ref{ha_all}). The width of the wings shows very high ejection velocity and we assume this to be related to nebular material ejected by the outbursting component. As the {\ha} profile had an appreciable FWHM of 1\,100 -- 1\,400\kms\ in addition, we conclude that the whole line was emitted mainly by the ejecta. In the period of our observations, the relative intensity of {\ha} increased at decreasing FWHM and a practically constant FWZI of 7\,000\kms\ (Fig.~\ref{ha_all}; and see below). However, {\ha} had very weak broad bumps on its wings at a velocity position of  $\pm(2\,300\div2\,400)$\kms\ as well. 
 The redshifted one is weaker and,  for this reason, less visible. 
The bumps are marked with arrows in the left panel of Fig.~\ref{ha_all}.
The line {\ha}  had similar bumps on its wings during the 2006 eruption as well, which were analysed by \citet{sk+08}. These authors concluded that these bumps represent satellite line components at the velocity position of about $\pm2\,400$\kms\ and are an indication of bipolar collimated outflow. On the other hand, the blueshifted bump seen during our observations was variable and was double peaked  on our last
spectrum, which means it is an individual component of the line (Fig.~\ref{ha_all}). This is why we are inclined to suppose that the emission bumps observed by us during the 2021 eruption of this system represent satellite components, and are an indication of bipolar outflow from the outbursting component as well. However, it should be noted that, as is seen in Fig.~2 from the work of \citet{sk+08}, the satellite components during the 2006 eruption were much more intense relative to the central emission of the line than during the 2021 eruption.


\begin{table}
\caption{Parameters of the central \ha\, emission. The flux is in units of $10^{-8}$ erg\,cm$^{-2}$\,s$^{-1}$, luminosity in (d/1.6kpc)$^{2}$ L$_{\sun}$, emission measure in $10^{61}$(d/1.6kpc)$^2$ cm$^{-3}$ and mass-loss rate in $10^{-5}$(d/1.6kpc)$^{3/2}$ M$_{\sun}$\,yr$^{-1}$. }
\label{tab:4}
\begin{center}
\begin{tabular}{@{}lcccc@{}}
\hline\hline
    \noalign{\smallskip}
Date   & $F$ & $L$ & $EM$ & $\dot M$ \\
    \noalign{\smallskip}
  \hline
     \noalign{\smallskip}
Aug~19 & 3.423 & 2700 & 3.74 & 5.0 \\
Aug~20 & 2.645 & 2100 & 2.89 & 4.3 \\
Aug~21 & 2.820 & 2200 & 3.08 & 4.3 \\
Aug~22 & 2.676 & 2100 & 2.92 & 4.0 \\
Aug~23 & 2.766 & 2200 & 3.02 & 4.0 \\
    \noalign{\smallskip}
\hline
    \noalign{\smallskip}

\end{tabular}
\end{center}
\end{table}


\begin{table*}
  \caption{Parameters of the satellite components. The flux is in units of $10^{-10}$ erg\,cm$^{-2}$\,s$^{-1}$, velocity and FWHM in \kms, linear angle $\theta$ in deg, and mass-loss rate in $10^{-6}$(d/1.6kpc)$^{3/2}$ M$_{\sun}$\,yr$^{-1}$.}
    \label{tab:5}
        \begin{center}
  \begin{tabular}{lccccccccccc}
  \hline\hline
    \noalign{\smallskip}
Date   & \multicolumn{5}{c}{Blue} & & \multicolumn{5}{c}{Red} \\
       \cline{2-6} \cline{8-12} \noalign{\smallskip}
         & $F$ & FWHM & $\upsilon / \cos i$ & $\theta$ & $\dot M$ & & $F$ & FWHM & $\upsilon / \cos i$ & $\theta$ & $\dot M$ \\
    \noalign{\smallskip}
  \hline
     \noalign{\smallskip}

Aug~19 & 3.320 & ~~880 & 3780 & 35.6 & 1.2 & & 1.706 & 670 & 3500 & 29.0 & 0.6 \\
Aug~20 & 2.846 & 1000 & 3720 & 41.2 & 1.2 & & 1.608 & 850 & 3460 & 37.5 & 0.8 \\
Aug~21 & 2.699 & 1040 & 3640 & 43.8 & 1.2 & & 1.352 & 780 & 3540 & 33.5 & 0.6 \\
Aug~22 & 2.243 & 1030 & 3620 & 43.7 & 1.0 & & 1.145 & 840 & 3500 & 36.7 & 0.6 \\
Aug~23 & 1.559 & ~~970 & 3620 & 40.9 & 0.8 & & 0.519 & 640 & 3610 & 26.7 & 0.3 \\

    \noalign{\smallskip}
\hline
    \noalign{\smallskip}

\end{tabular}
\end{center}

\end{table*}

To obtain the parameters of the stellar wind and bipolar outflow, we analysed the \ha\, profile by means of approximation with different functions: because of its asymmetry, the central emission together with the wings was approximated with a sum of Gaussian and Lorentzian and each bump was approximated with one Gaussian (Fig.~\ref{ha_all}). For approximation of the bumps, we adopted FWZI~=~2\,FWHM. The parameters of the central emission and satellite components obtained with this procedure are listed in Tables~\ref{tab:4} and \ref{tab:5}. 
The error of the equivalent width of the central emission 
 is determined primarily by the uncertainty 
 in the approximation of 
 the continuum level and is less than 1.5\%. The error of the flux, luminosity, and emission measure of the central emission is less than 5\%. The bumps are weak broad emission components and it is difficult to determine their extension precisely.
The error of their equivalent width results from their approximation and the uncertainty 
 in the approximation of
 the continuum level and is not more than 26\% for all spectra. The errors on their FWHM and velocity position result mainly from the approximation; the former is 3\% and the latter is 2\%. 
The error on the energetic flux of the satellite components is 26\%. 
        
A number of the analyses presented here show that the ejection of mass during the outburst of \rsoph\ is not spherical and is closer to bipolar \citep{o'brien+06, booth+16, montez+22}. \citet{booth+16} modelled the circumstellar medium in \rsoph\ in quiescence and during outburst. These authors obtained a bipolar structure of the wind of the outbursting component resulting from the interaction of a spherical nova outburst with the accretion disc. As is seen from their Fig.~3, the wind propagates in a linear angle of about 120\degr. For our calculations, we accept a model of a bipolar wind propagating at a linear angle of 120\degr\ as well. We calculated the mass-loss rate $\dot{M}$ from the \ha\, flux supposing that the wind has a constant velocity $\upsilon$ and using the nebular approach of \citet{VN94}. The energy flux of a bipolar nebula with a solid angle $\Omega$ and a radius $r$ at a distance $d$ is 

\begin{equation}
        F=\frac{\Omega}{4\pi} \frac{\alpha h\nu}{d^2}\int_r n_{\rm e}n^{+}r^2\,{\rm d}r \,,
\end{equation}  
 
\noindent
where $\alpha$ is the recombination coefficient at the upper level of  \ha\, transition. The particle density in the wind is 

\begin{equation} \label{eq2}
        n(r)=\frac{\dot{M}}{\Omega r^2 \mu m_{\rm H}\upsilon} \,\,,
\end{equation}

\noindent
where $\mu\!=\!1.4$ \citep{NV87} is the parameter determining the mean molecular weight $\mu m_{\rm H}$ in the wind. For the velocity of the wind, we accepted $\upsilon\!=\!3\,500\pm250$\kms\  which is equal to the  
{\ half width at zero intensity (HWZI)}
 of the line throughout our observations. The data in Table~\ref{tab:4} show the $EM$ of \ha\ is 30\%--50\% of the $EM$ based on the nebular continuum for August 19 and 23, which means that a significant part of the nebular emission of the system is produced by the wind. We then assumed that the dominant state of ionisation of helium in the wind is also {\heii} and the electron temperature is 17\,000~K. We used a recombination coefficient for a temperature of 17\,000~K and a density of 10$^{11}$~cm$^{-3}$ \citep{SH}, which is very close to the mean density in the wind. For the inner boundary of the wind, we took the  radius of the outbursting component calculated at the end of Section 3. 
The outer boundary of integration was determined in the following way. After the beginning of the outburst, the ejected material creates an expanding lobe inside the wind of the giant \citep{GW87,NW93,bisikalo+06}. The expansion velocity decreases with time because the expansion is slowed by the wind of the giant, and reaches a final value of 

\begin{equation}
        \upsilon_{\rm shell}\!=\!\upsilon \frac{1+\sqrt{m\omega}}{\omega+\sqrt{m\omega}}\,\,,
\end{equation}

\noindent
where $m\!=\!\dot{M} / \dot{M_0}$, $\omega\!=\!\upsilon / \upsilon_0$, and $\upsilon_0$, $\dot{M_0}$, $\upsilon,$ and $\dot{M}$ are the wind velocities and mass-loss rates of the giant and outbursting component, respectively \citep{GW87}. Using the parameters of the wind from the giant $\upsilon_0 \!=\! 20$~\kms and  $\dot{M_0}\!=\!5\,10^{-7}$~M$_{\sun}$\,yr$^{-1}$ \citep{booth+16} and the parameters that we obtained for the outbursting component $\upsilon \!=\! 3500$~\kms and  $\dot{M}\!=\!5\,10^{-5}$~M$_{\sun}$\,yr$^{-1}$, we derived $\upsilon_{\rm shell}\!=\!1520$~\kms. For 15 days after the beginning of the outburst, the lobe, which is filled by this newly appearing wind and is expanding with this latter velocity, should reach a size of 13.2~au. However, as our observations were made at a very early stage of the outburst, we suppose that the real expansion velocity is closer to the observed wind velocity than to $\upsilon_{\rm shell}$ and therefore we used the observed velocity. For an expansion for 15 days, we obtained a size of the lobe of 30.3~au. This size was used as an outer boundary of the wind.
 The parameters of the wind are listed in Table~\ref{tab:4} for each date of observation. 
 The error of the mass-loss rate is 7\%.
The change of the mass-loss rate from $(5.0\pm0.4)\,10^{-5}$(d/1.6kpc)$^{3/2}$ M$_{\sun}$\,yr$^{-1}$ to $(4.0\pm0.3)\,10^{-5}$(d/1.6kpc)$^{3/2}$ M$_{\sun}$\,yr$^{-1}$ is greater than this latter error and we conclude that it decreased during our observations.
We hypothesise that the rate decreased more quickly but our result is influenced by the increase in the contribution of the other parts of the circumbinary nebula in the central emission of \ha\ when the system returns to its quiescent state. The decrease in the wind contribution to the central emission is indicated by the diminution of its FWHM. This is why we consider our result as an upper limit of the mass-loss rate. \citet{sk+08} obtained a mass-loss rate of the outbursting component of $(1-2)\,$10$^{-4}$(d/1.6kpc)$^{3/2}$ M$_{\sun}$\,yr$^{-1}$ for day 1.38, which is close to the light maximum of the 2006 eruption of {\rsoph}. The mass-loss rate that we obtain here for days 11--15 of the 2021 eruption is less than $(4-5)\,$10$^{-5}$(d/1.6kpc)$^{3/2}$ M$_{\sun}$\,yr$^{-1}$. 
 
The next task of our investigation is to calculate the mass-loss rate from the bipolar streams of the outbursting component, which are indicated by the satellite components. We used the same nebular approach, supposing that the outflowing material propagates with a constant velocity in two spherical sectors with a small linear angle. The linear angle $\theta$ and solid angle $\Omega$ of a spherical sector were calculated using the parameters of the satellite components and orbit inclination, adopting the approach of \citet{sk+09}. We used an orbit inclination of 50\degr\ according to \citet{br+09}. We
also used the same electron temperature and boundaries of integration as for the wind. The parameters of the two streams are listed in Table~\ref{tab:5} for each date of observation. The mean velocities of the satellite components of $\upsilon_{b}\!=\!-2\,360\pm40$\kms\ for the blueshifted component and $\upsilon_{r}\!=\!2\,260\pm30$\kms\ for the redshifted component are very close to their velocity during the 2006 eruption \citep{sk+08}. The linear angle $\theta$ shows that the streams are not highly collimated. 
 The error of the mass-loss rate of the compact object through the streams is 26\%.
The change of the mass-loss rate slightly exceeds its error and we conclude that it did not vary during our observations, having a mean value of $\dot{M}_{b}\!=\!(1.1\pm0.3)\,$10$^{-6}$ (d/1.6kpc)$^{3/2}$ M$_{\sun}$\,yr$^{-1}$ for the approaching stream and $\dot{M}_{r}\!=\!(0.6\pm0.2)\,$10$^{-6}$ (d/1.6kpc)$^{3/2}$ M$_{\sun}$\,yr$^{-1}$ for the receding one.
It is seen from Tables~\ref{tab:4} and \ref{tab:5} that the mass-loss rate from the streams is a small part of the total mass-loss rate (wind + streams) for the outbursting object. 

The density in an outflow with a constant velocity decreases with the radius as $r^{-2}$ (Eq.~\ref{eq2}). However, we can obtain one tentative estimate of the size of the high-velocity portion of the streams emitting satellite components if we consider them as spherical sectors with a constant density. In this case, we can obtain their volume from their emission measure. The density in the streams is very close to the density in the wind and we assumed the same mean density of 10$^{11}$~cm$^{-3}$. The mean emission measure and solid angle of the approaching stream of 2.77\,10$^{59}$~cm$^{-3}$ and 0.4010~sr lead to a radius of the spherical sector of $r_{b}\!=\!(85\pm8)$(d/1.6kpc)\,R$_{\sun}$ if the error of the emission measure is 26\%.  The mean emission measure and solid angle of the receding stream of 1.38\,10$^{59}$~cm$^{-3}$ and 0.2584~sr lead to a radius of the spherical sector of $r_{r}\!=\!(78\pm7)$(d/1.6kpc)\,R$_{\sun}$. These radii are tentative estimates of the size of the streams. 
 
 \section{Discussion} 

For the first and last day of our observation, August 19 and 23, we find the rate of recombination in the nebula of {\rsoph} to be much greater than the Lyman luminosity of the ionising star. To balance the rate of recombination, the Lyman luminosity should be $L_{\rm ph}\gid1.74\,10^{49}$~phot\,s$^{-1}$ and $L_{\rm ph}\gid1.03\,10^{49}$~phot\,s$^{-1}$ for these two days. According to \citet{sk15b}, the temperature of the outbursting object increases to more than 100\,000~K after the light maximum of the 2006 eruption of {\rsoph}. We accepted a temperature of 150\,000~K and radii  of 0.9~R$_{\sun}$ and 0.7~R$_{\sun}$ for August 19 and 23, and obtained Lyman luminosities of 2.12\,10$^{49}$~phot\,s$^{-1}$ and 1.28\,10$^{49}$~phot\,s$^{-1}$, respectively. These temperature and radii give enormous bolometric luminosities of 3.7\,10$^{5}$~L$_{\sun}$ and 2.2\,10$^{5}$~L$_{\sun}$, which exceed the Eddington luminosity of the hot component in {\rsoph} by factors of 8 and 5, respectively. From his multi-wavelength modelling of the continuum energy distribution, \citet{sk15a} obtained a temperature of 110\,000~K and a radius of 2.1~R$_{\sun}$ at day 7.3 after the beginning of the 2006 outburst, which lead to an equally enormous bolometric luminosity of 5.8\,10$^{5}$~L$_{\sun}$, exceeding the Eddington limit by a factor of more than 13. At day 19.5, the bolometric luminosity exceeded the Eddington limit by a factor of more than 4. These data propose a similar behaviour of the outbursting component during the two eruptions. Immediately after the maximal light of both eruptions, the emission measure of the circumbinary nebula was ~10$^{62}$~cm$^{-3}$ \citep[][this paper]{sk15a}. One remarkable feature of the {\rsoph}  system is that during the 2021 eruption, about 30\%--50\% of the nebular emission belongs to the high-velocity wind. The luminosity of the wind in \ha\ reaches ~2900~L$_{\sun}$ during the 2006 eruption \citep{sk+08} and less than 2700~L$_{\sun}$ during the 2021 one (result from the present paper). 

\section{Conclusion} 

We present results of high-resolution \ha\ observations and simultaneous $U$, $B$, $V$, $R_{\rm C}$, and $I_{\rm C}$ photometry from the AAVSO database carried out soon after the light maximum of the 2021 outburst of the symbiotic nova {\rsoph}. We analysed the continuum energy distribution of the system at the first and last day of our observation, namely August 19 (day 11 of the outburst) and August 23 (day 15). Our results show that the effective temperature of the outbursting component was $T_{\rm eff}\!=\!15\,000\pm1\,000$~K and the electron temperature of the nebula $T_{\rm e}\!=\!17\,000\pm3000$~K for the whole period of observation. The effective radius of the observed photosphere (pseudophotosphere) was $R_{\rm eff}\!=\!(13.3\pm2.0)$(d/1.6kpc)~R$_{\sun}$ and the emission measure of the nebula $EM\!=\!(9.50\pm0.59)$\,10$^{61}$(d/1.6kpc)$^{2}$~cm$^{-3}$ for August 19 and $R_{\rm eff}\!=\!(10.3\pm1.6)$(d/1.6kpc)~R$_{\sun}$ and $EM\!=\!(5.60\pm0.35)$\,10$^{61}$(d/1.6kpc)$^{2}$~cm$^{-3}$ for August 23. These data show that the Lyman photon luminosity of this pseudophotosphere is insufficient to balance the rate of recombination in the nebula, which, in turn, means we observed a disc-shaped warm shell occulting the central hot object. This shell could appear as the result of a collision between the ejected material and an accretion disc. The hot object ionises the nebula up and down and produces the observed emission measure. To provide this emission measure, its bolometric luminosity should exceed several times its Eddington limit. 

We analysed the \ha\ profile with the aim being to study the outflow structure. Throughout the period of observation, the \ha\ was very intensive broad emission line with an FWHM of 1\,100--1\,400\kms\ and FWZI of 7\,000\kms, possessing weak broad bumps at its wings with mean velocities of $\upsilon_{b}\!=\!-2\,360\pm40$\kms\ and $\upsilon_{r}\!=\!2\,260\pm30$\kms. We conclude that most of the  central emission is produced by the ejected material from the outbursting component (stellar wind) and the bumps are satellite components indicating bipolar outflow. We also find that about 30\%--50\% of the nebular emission of the system belongs to the wind, the \ha\ luminosity of which was less than 2700~L$_{\sun}$. We determined the mass-loss rate of the outbursting component from those of the wind and the streams. 
The mass-loss rate from the streams is a small part of the total rate, whose upper limit decreased from $(5.2\pm0.4)$\,10$^{-5}$ (d/1.6kpc)$^{3/2}$ M$_{\sun}$\,yr$^{-1}$ to $(4.1\pm0.3)$\,10$^{-5}$ (d/1.6kpc)$^{3/2}$ M$_{\sun}$\,yr$^{-1}$ during our observations.
Our results suggest the streams were not highly collimated, and we obtained one tentative estimate of their size, which amounts to about 80(d/1.6kpc)\,R$_{\sun}$.

\begin{acknowledgements}

The authors are grateful to an anonymous referee for his critical remarks which led to an improvement of the manuscript.
We are grateful to Pierre Dubreuil (PDL), Joan Guarro (JGF), Keith Shank (KSH) and Francois Teyssier (FTE) for their spectroscopic observations as well.
We acknowledge with thanks the variable star observations from the AAVSO International Database contributed by observers worldwide and used in this research.
The authors gratefully acknowledge observing grant support from the Institute of Astronomy and National Astronomical Observatory, Bulgarian Academy of Sciences.

This work was partially supported by Bulgarian National Science Fund of the Ministry of Education and Science under grants KP-06-Russia/2-2020 and DN~18-13/2017. DM acknowledges partial support by Shumen University Science Fund. PK was supported by RFBR grant 20-52-18015.

\end{acknowledgements}

\listofobjects


\begin{thebibliography}{}


\bibitem[Banerjee et al.(2009)]{ban+}
        Banerjee, D. P. K., Das, R. K., \& Ashok, N. M. 2009, MNRAS, 399, 357

\bibitem[Bessell(1979)]{bessell79}
        Bessell, M. S. 1979, PASP, 91, 589

\bibitem[Bisikalo et al.(2006)]{bisikalo+06}
        Bisikalo, D. V., Boyarchuk, A. A., Kilpio, E. Yu., Tomov, N. A., \& Tomova, M. T. 2006, ARep, 50, 772
        
\bibitem[Bode(1987)]{bode87}
        Bode, M. 1987, \rsoph\ (1985) and the Recurrent Nova Phenomenon, VNU Science Press, p. 241

\bibitem[Booth et al.(2016)]{booth+16}
     Booth, R. A., Mohamed, S., \& Podsiadlowski, Ph. 
        2016, MNRAS, 457, 822

\bibitem[Brandi et al.(2009)]{br+09}
        Brandi, E., Quiroga, C., Mikolajewska, J., Ferrer, O. E., \& Garcia, L. G. 
        2009, A\&A, 497, 815
        
\bibitem[Cardelli et al.(1989)]{Cardelli}
        Cardelli, J. A., Clayton, G. C., \& Mathis, J. S. 1989, ApJ, 345, 245

\bibitem[Fekel et al.(2000)]{f+00}
        Fekel, F. C., Joyce, R. R., Hinkle, K. H., \& Skrutskie, M.
         2000, AJ, 119, 1375

\bibitem[Ferland(1980)]{ferland80}
        Ferland, G. J. 1980, PASP, 92, 596

\bibitem[Geary(2021)]{geary21}
        Geary, K. 2021, AAVSO Alert Notice 752 (20210809)

\bibitem[Girard \& Willson(1987)]{GW87}
        Girard, T., \& Willson, L. A. 1987, A\&A, 183, 247

\bibitem[Iijima(2009)]{Iijima09}
        Iijima, T. 2009, A\&A, 505, 287
        
\bibitem[Magano(2017)]{magano+17}
        Magano, D. M. N., Vilas Boas, J. M. A., \& Martins, C. J.       A. P. 2017, Phys. Rev. D, 96, 083012

\bibitem[Mikolajewska et al.(2021)]{Mik+21}
        Mikolajewska, J., Aydi, E., Buckley, D., Galan, C., \& Orio, M. 2021, ATel 14852

\bibitem[Montez et al.(2022)]{montez+22} 
        Montez, R., Luna, G. J. M., Mukai, K., Sokoloski, J. L.,        \& Kastner, J. H. 2022, ApJ, 926, 100

\bibitem[Munari \& Valisa(2021a)]{MV21a}
        Munari, U., \& Valisa, P. 2021a, ATel 14840
        
\bibitem[Munari \& Valisa(2021b)]{MV21b}
        Munari, U., \& Valisa, P. 2021b, arXiv:2109.01101
        
\bibitem[N\"ussbaumer \& Vogel(1987)]{NV87}
        N\"ussbaumer, H., \& Vogel, M. 1987, A\&A, 182, 51

\bibitem[N\"ussbaumer \& Walder(1993)]{NW93}
        N\"ussbaumer, H., \& Walder, R. 1993, A\&A, 278, 209

\bibitem[O'Brien et al.(2006)]{o'brien+06}
        O'Brien, T. J., Bode, M. F., Porcas, R. W., et al.
        2006, Nature, 442, 279

\bibitem[Pottasch(1984)]{Pottash} 
        Pottasch, S. R. 1984, Planetary nebulae.  Reidel, Dordrecht, p. 93

\bibitem[Skopal(2015a)]{sk15a}
        Skopal, A. 2015a, New Astron., 36, 128

\bibitem[Skopal(2015b)]{sk15b}
        Skopal, A. 2015b, New Astron., 36, 139

\bibitem[Skopal et al.(2008)]{sk+08}
     Skopal, A., Pribulla, T., Buil, Ch., Vittone, A., \&       Errico, L.
     2008, in ASP Conf. Ser. 401, RS~Ophiuchi 2006 and Recurrent        Nova Phenomenon, 
     ed. A. Evans, M. F. Bode, T. J. O'Brien, \& M. J. Darnley, 227

\bibitem[Skopal et al.(2009)]{sk+09}
         Skopal, A., Pribulla, T., Budaj, J., et al. 
         2009, ApJ, 690, 1222

\bibitem[Skopal et al.(2013)]{sk+13}
         Skopal, A., Tomov, N. A., Tomova, M. T. 2013, A\&A, 551, L10

\bibitem[Sokoloski et al.(2006)]{sokoloski+06}
        Sokoloski, J.L.,  Luna, G. J. M., Mukai, K., Kenyon, S. J.      2006, Nature, 442, 276

\bibitem[Storey \& Hummer(1995)]{SH}
        Storey, P. J., \& Hummer, D. G., 1995, MNRAS, 272, 41

\bibitem[Taguchi et al.(2021)]{taguchi+21}
        Taguchi, K., Maehara, H., Isogai, K., Tampo, Y., \& Ito, J. 2021, ATel 14858

\bibitem[Teyssier(2019)]{teyssier19}
        Teyssier, F. 2019, Contrib. of the Astron. Obs. Skalnate Pleso, 49, 217
        
\bibitem[Tomov et al.(2007)]{TTB07}
        Tomov, N. A., Tomova, M. T., \& Bisikalo, D. V. 2007,   MNRAS, 376, L16 

\bibitem[Tomov \& Kolev(1997)]{TK97}
        Tomov, T., \& Kolev, D. 1997, A\&AS, 122, 43

\bibitem[Tomov et al.(2000)]{toma00}
        Tomov, T., Munari, U., \& Marrese, P. 2000, A\&A, 354, L25

\bibitem[Vogel \& N\"ussbaumer(1994)]{VN94}
        Vogel, M., \& Nussbaumer, H. 1994, A\&A, 284, 145
        
\bibitem[Zamanov et al.(2018)]{zam+18} 
        Zamanov, R., Boeva, S., Latev, G., et al. 2018, MNRAS, 480, 1363

\bibitem[Zamanov et al.(2022)]{zam+22} 
        Zamanov, R. K.,Stoyanov, K. A., Nikolov, Y. M., et al. 
        2022, BlgAJ, 37, 24

\end{thebibliography}
\end{document}